**TITLE**

Effects of pristine and photoaged tire wear particles and their leachable additives on key nitrogen removal processes and nitrous oxide accumulation in estuarine sediments

Jinyu Ye [a,b*], Yuan Gao [a,b], Huan Gao [c], Qingqing Zhao [a,b], Minjie Zhou [d], Xiangdong Xue [a,b*], Meng Shi [e]

a. School of Civil Engineering and Architecture, Zhejiang University of Science and Technology, Hangzhou, Zhejiang, 310023, China

b. Zhejiang-Singapore Joint Laboratory for Urban Renewal and Future City, Hangzhou, 310023, China

c. Key Laboratory of Pesticide & Chemical Biology of Ministry of Education, College of Chemistry, Central China Normal University, Wuhan, 430079, China

d. Pingyang County Aojiang River Basin Water Conservancy Project Management Center, Wenzhou, Zhejiang,325401, China

e. Center for Energy Resources Engineering, Department of Chemical and Biochemical Engineering, Technical University of Denmark, Kongens Lyngby, 2800, Denmark

\* Corresponding author: Xiangdong Xue, School of Civil Engineering and Architecture, Zhejiang University of Science and Technology, No.318 Liuhe Road, Zhejiang, 310023, China; E-mail: water21cn_xxd@163.com

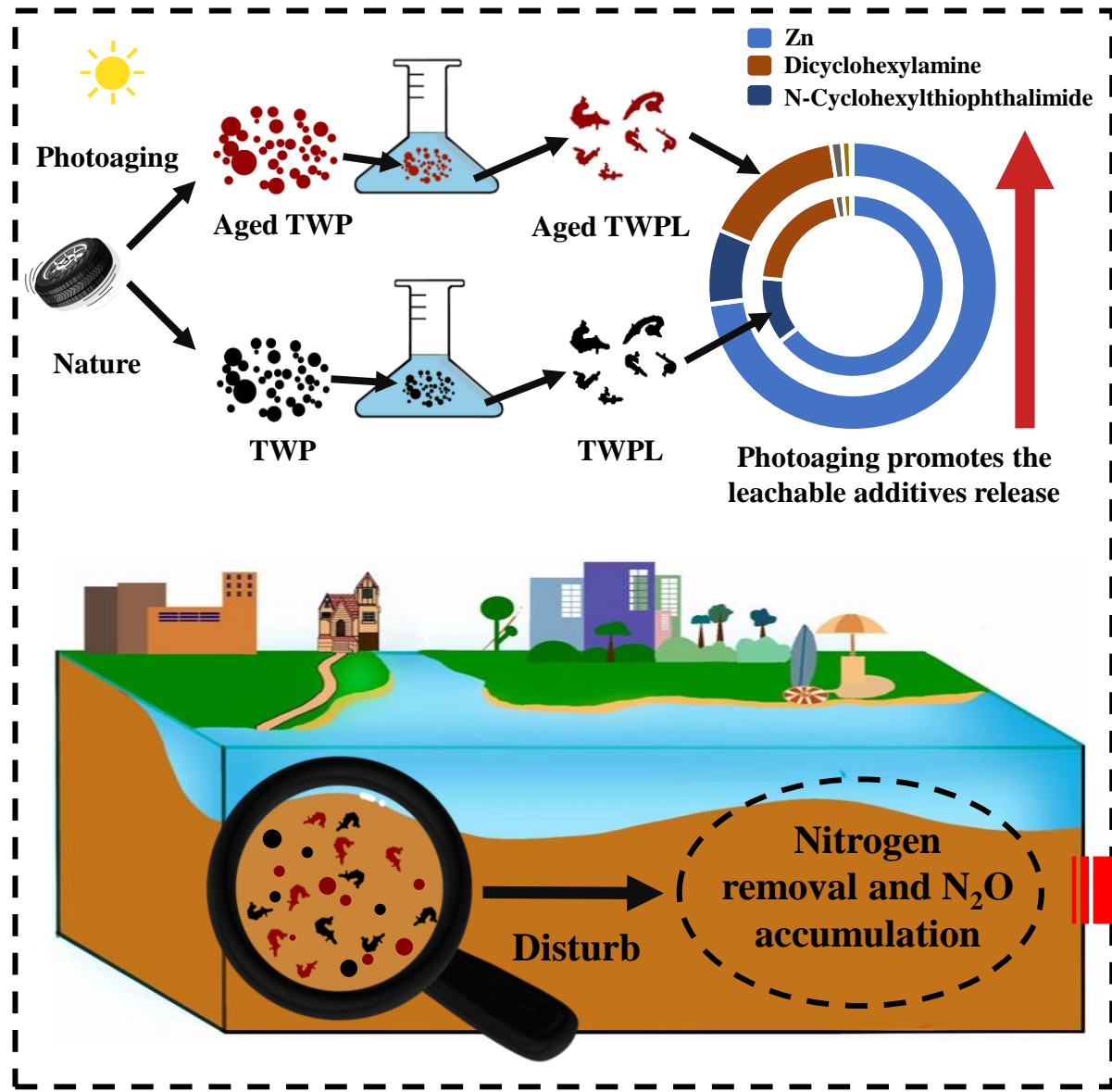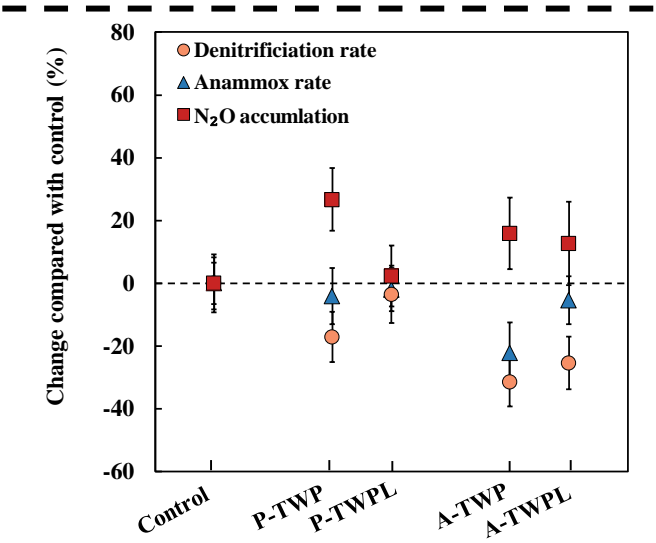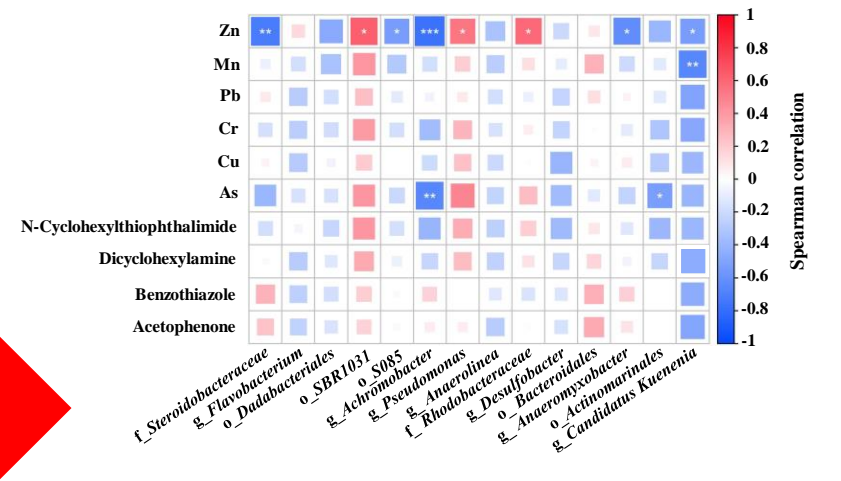

**Abstracts:** Global estuaries and coastal regions, acting as critical interfaces for mitigating nitrogen flux to marine, concurrently contend with contamination from tire wear particles (TWPs). However, the effects of pristine and photoaged TWP (P-TWP and A-TWP) and their leachates (P-TWPL and A-TWPL) on key nitrogen removal processes in estuarine sediments remain unclear. This study explored the responses of denitrification rate, anammox rate, and nitrous oxide ($N_2O$) accumulation to P-TWP, A-TWP, P-TWPL, and A-TWPL exposures in estuarine sediments, and assessed the potential biotoxic substances in TWPL. Results indicate that P-TWP inhibited the denitrification rate and increased $N_2O$ accumulation without significantly impacting the anammox rate. A-TWP intensified the denitrification rate inhibition by further reducing *narG* gene abundance and NAR activity, and also decreased the *hzo* gene abundance, HZO activity, and *Candidatus Kuenenia* abundance, thereby slowing the anammox rate. $N_2O$ accumulation was lower after A-TWP exposure than P-TWP, with the NIR/NOS and NOR/NOS activity ratios closely associated with $N_2O$ accumulation. Batch experiments indicated that photoaging promoted Zn release from TWPL, significantly contributing to the inhibited denitrification rate and increased $N_2O$ accumulation by TWP. In addition, TWP drives changes in microbial community structure through released additives, with the abundance of DNB and AnAOB closely linked to the Zn, Mn, and As concentrations in TWPL. This study offers insights into assessing the environmental risks of TWPs in estuarine ecosystems.

# 1. Introduction

Tire wear particles (TWPs), produced from the friction between tire treads and road surfaces, are considered a typical microplastic (MP) pollution in the environment (Xiao et al., 2024). Estimates suggest that the total annual global release of TWPs is approximately 5918 kt, corresponding to an annual per capita release of 0.81 kg (Kole et al., 2017). Road simulators have determined that most TWPs have a diameter smaller than 200 um (Kreider et al., 2010). These particles can enter the soil, the water bodies, and the atmosphere due to atmospheric transport and surface runoff, potentially entering the human body through the food chain (Hua and Wang, 2023). Therefore, the environmental fate and ecological risk of TWPs have attracted widespread attention.

Estuarine regions, typically highly developed, exhibit high pollution levels of TWPs due to advanced transportation networks and high vehicle density (Barber et al., 2024). The concentration of MPs in estuarine sediments such as Hangzhou Bay and Charleston Harbor ranged from 84.3 particles/kg - 413.8 particles/kg, with over 70% identified as TWPs (Gray et al., 2018; Wang et al., 2020). As interface zones between land and sea, estuarine ecosystems face severe environmental challenges due to eutrophication exacerbated by increased anthropogenic nitrogen loading (Liang et al., 2024). Microbially driven denitrification and anammox processes are central to nitrogen removal in estuaries (Wu et al., 2021). Denitrifying bacteria (DNB) reduce nitrate ($NO_3^-$) sequentially to nitrite ($NO_2^-$), nitric oxide (NO), nitrous oxide ($N_2O$), and nitrogen gas ($N_2$), while anammox bacteria (AnAOB) use $NO_2^-$ produced during denitrification as a receptor to convert ammonia ($NH_4^+$) to $N_2$ (Wu et al., 2022). Key enzymes involved in these two processes include $NO_3^-$ reductase (NAR, encoded by *narG*), $NO_2^-$ reductase (NIR, encoded by *nirS* and *nirK*), NO reductase (NOR, encoded by *norB*), $N_2O$ reductase (NOS, encoded by *nosZ*) and hydrazine synthase (HZO, encoded by *hzo*) (Xu et al., 2023). A recent study explored the response of microbial community structure and diversity in sediments to TWPs at the genetic level and found that TWPs generally reduce the abundance of nitrogen-transforming microbial communities (Ding et al., 2022). However, a decrease in microbial abundance does not necessarily imply a reduction in nitrogen transformation activity (Ali et al., 2020), and there is currently no evidence that TWPs disrupt critical nitrogen transformation steps in estuarine sediments.

Nitrogen removal in estuaries is accompanied by the release of $N_2O$, a greenhouse gas primarily produced through incomplete denitrification (Su et al., 2021). The global warming potential of $N_2O$ is over 300 times that of carbon dioxide ($CO_2$) and accounts for about 6% of global greenhouse effects (Chen et al., 2022; Ye et al., 2024). Estuarine and coastal regions contribute 12% - 18% of global $N_2O$ emissions (Su et al., 2021). Several studies have pointed out that MPs might influence $N_2O$ production during denitrification. For example, 0.4% (w/w) of microfibers reduced $N_2O$ emissions during soil denitrification (Rillig et al., 2021), while 1% (w/w) of polyethylene MPs enhanced $N_2O$ emissions during sediment denitrification (Chen et al., 2022). Given the potential impact of TWPs on nitrogen removal and $N_2O$ emissions in estuarine sediments, there is an urgent need to elucidate the responses of key microbial nitrogen removal processes, such as denitrification and anammox, to TWP pollution in estuarine ecosystems.

In addition to altering sediments' hydrophobicity and hydrological characteristics through particle-induced effects, TWPs also impact the structure and function of microbial communities in sediments by releasing leachable additives (Liu et al., 2022b). Compared to thermoplastic MPs, TWPs generally have greater biotoxicity due to containing over 300 types of heavy metals and organic additives, most of which are toxic to microbes (Wik and Dave, 2009). Recent studies have revealed various additives discovered in the pore water of sediments exposed to TWPs, which can explain over 90% of the changes in community structure (Ding et al., 2022). Aging is an inevitable process for TWPs in natural environments and involves photoaging, thermal aging, biodegradation, and mechanical fragmentation, with photoaging recognized as the dominant form (Ding et al., 2024). Previous studies have demonstrated that photoaged TWP (A-TWP) exhibits higher toxicity towards *Bacillus subtilis* and *Halioglobus lutimaris* compared to pristine TWP (P-TWP), partly due to increased toxicity in TWP leachate (TWPL) caused by photoaging (Liu et al., 2022b). The $Zn^{2+}$ concentration in the A-TWPL was 24 times higher than that in the P-TWPL, resulting in a 3.3% increase in the mortality rate of *Daphnia magna* (Lu et al., 2021). However, existing TWP toxicological studies mainly focus on typical model organisms, with limited research comparing the effects of P-TWP and A-TWP on key nitrogen removal processes and $N_2O$ accumulation in estuarine sedments, particularly the toxic contribution of their leachable additives.

Based on the above, this study hypothesizes that photoaging inevitably alters the impact of TWPs on key nitrogen removal processes and $N_2O$ accumulation in estuarine sediments, with variations in TWPL composition and concentration partly driving these changes. The objectives of this study include: 1) Investigate the effects of P-TWP and A-TWP on denitrification rate, anammox rate, and $N_2O$ accumulation in estuarine sediments; 2) Explore the response mechanisms of key nitrogen removal processes and $N_2O$ accumulation to P-TWP and A-TWP by analyzing nitrogen transformation enzyme activities, gene abundances, and microbial community structure; 3) Clarify the biotoxic contribution of TWPL to TWP and identify potential toxic substances in TWPL.

## 2. Materials and methods

### 2.1 Preparation of P-TWP, A-TWP, P-TWPL and A-TWPL

Waste tires were shredded into approximately 30 mm particles using tire recycling equipment, then further milled into pristine P-TWP with an average size below 200 μm using a cross-beater mill (SK300, Retsch, Haan, Germany). To produce A-TWPs, P-TWP was exposed to 1000 W xenon lamp irradiation (wavelength 500 - 2000 nm) at 25±1°C for 40 days in a photochemical reactor (BL-GHX-V, Bilong, Shanghai, China). The particle size distribution and surface morphology of P-TWP and A-TWP were confirmed using dynamic light scattering (DLS, Zetasizer Nano ZS90, Malvern Instruments, Worcestershire, UK) and scanning electron microscopy (SEM, Regulus 8100, Hitachi, Tokyo, Japan), as shown in Fig.S1.

To prepare P-TWPL or A-TWPL, 0.5g P-TWP or A-TWP were mixed with 50 ml of estuarine overlying water and agitated at 25±1°C and 200 rpm for 40 days using an orbital shaker (ZD-9550, Beyotime, Shanghai, China). After agitation, the particles were removed using a 0.22 μm cellulose acetate filter (Thermo Fisher Scientific, OH,

USA) to obtain P-TWPL or A-TWPL.

**2.2 Sample collection and experimental design**

Sediment and overlying water were collected from Hangzhou Bay, Zhejiang Province, China, at coordinates 30.4067N, 121.2084E (Fig. S2). Complete sediment cores (0 - 20 cm) and overlying water were collected using a portable corer (Rigo Co., Saitama, Japan) and polymethyl methacrylate sampler (UWITEC, Mondsee, Austria), respectively. All samples were preserved at 4°C and promptly transported to the laboratory. Before the experiments, overlying water was filtered through a 0.22 um cellulose acetate filter, while sediments were thoroughly homogenized and then passed through a 2 mm sieve to remove larger particles. The physicochemical properties of the sediments and overlying water are shown in Table S1.

Previous studies have reported that TWPs account for 0.5% of the total sediment weight in the seine river basin (Unice et al., 2013). Therefore, this study set the TWP concentration at 0.5% (w/w) to reflect environmental concentrations. Additionally, considering that TWPs release various additives in aquatic environments, the impact of leachable additives is equivalent to 0.5% (w/w) of TWPs was evaluated. Homogenized sediments were mixed with TWP or TWPL and dispensed into sterilized sediment cores with a diameter of 10 cm and a height of 50 cm (Fig. S3). Overlying water was maintained at a level 1 cm above the sediment. The cores were wrapped in aluminum foil to exclude light and incubated at $25 \pm 1$°C for 40 days, with periodic adjustments to the overlying water level every three days. The experiment consisted of five treatment groups: control, P-TWP, A-TWP, P-TWPL, and A-TWPL, each conducted in triplicate.

**2.4 Heavy metals and organic compounds analysis**

Sediment samples (5 g) were centrifuged at 10,000 rpm and 4°C for 10 minutes to separate the pore water. Before analysis, the pore water was filtered through a 0.22 μm cellulose acetate filter. Concentrations of heavy metals and organic compounds were detected using inductively coupled plasma emission spectroscopy (ICP-OES, Perkin Elmer, CT, USA) and gas chromatography-tandem mass spectrometry (GC-MS/MS, Agilent, CA, USA) in multiple reaction monitoring modes.

**2.3 Denitrification, anammox, and $N_2O$ production rates measurement**

After 40 days of incubation, sediment cores were sealed, and $^{15}N$-labeled nitrate ($NaNO_3$) was added via a syringe interface to each column's overlying water to a final concentration of approximately 5 mg N/L. Then, the cores were incubated for 24 hours in darkness. At the start and after 24 h of incubation, overlying water was sampled into 12 mL exetainers (Labco Exetainer, High Wycombe, UK). Each exetainer was supplemented with 200 μL of saturated $ZnCl_2$ solution to inhibit microbial activity for subsequent isotopic analysis and quantification of dissolved $N_2$ and $N_2O$. Concentrations of $^{29}N_2$ and $^{30}N_2$ were analyzed using a membrane inlet mass spectrometer (MIMS, Bay Instruments, Maryland, USA).

For the analysis of $N_2O$ isotopic ratios and concentrations, 4 mL of analytical-grade He was injected into exetainers filled with overlying water at atmospheric pressure using a syringe and a two-way valve. Then, the exetainers were vigorously shaken and inverted to allow $N_2O$ to equilibrate between the water and the headspace.

The ratios of $^{14}NO_3^-$ to $^{15}NO_3^-$ (r14) and the $N_2O$ concentrations in the headspace were measured using an isotope ratio mass spectrometer (IRMS, Delta V Advantage, Thermo Fisher Scientific, MA, USA) and a gas chromatograph (GC-2014, Shimadzu, Kyoto, Japan), respectively. Denitrification, anammox, and $N_2O$ accumulation rates were calculated based on previous methods (Wu et al., 2022).

**2.5 Enzyme activities, gene abundance, and microbial community structure assay**

Sediment samples (5 g) were washed 3 times in 100 mM phosphate-buffered saline (PBS, pH=7.8) and resuspended at 4°C. The resuspension was sonicated at 20 kHz for 10 min, followed by centrifugation at 12,000 rpm for 10 minutes to collect the supernatant for enzyme activity measurement. The activities of NAR, NIR, NOR, and NOS were determined based on previous studies (Ye et al., 2021). HZO activity was analyzed using an enzyme-linked immunosorbent assay (ELISA) kit (FIYA Bio, Shenzhen, China) based on specific interactions between antibodies and the corresponding antigens (Xu et al., 2023).

Total genomic DNA was extracted from sediment samples using the FASTDNA® Spin Kit (MP Biomedicals, OH, USA). DNA concentration and quality were determined using a NanoDrop® spectrophotometer (Nanodrop, DE, USA). The abundance of *narG*, *nirK*, *nirS*, *norB*, *nosZ*, and *hzo* genes was quantified using Real-time Quantitative PCR (RT-qPCR). All RT-qPCR reactions were performed on a CFX Connect real-time system (Bio-Rad, CA, USA) using forward and reverse primers (Table S2). The microbial community structure was analyzed through the V3-V4 regions of the 16S rRNA gene using primers 338F and 806R (Ye et al., 2023). High-throughput sequencing was performed using an Illumina MiSeq platform (Annoroad Gene Technology, Beijing, China). Sequence analyses were carried out using the analytical software in Majorbio I-Sanger Cloud (v2.0, www.i-sanger. com).

**2.6 Released Zn and As impact experiment setups**

The theoretical concentrations of total Zn and As in sediments after exposure to P-TWP or A-TWP were estimated based on the concentrations of these elements detected in the corresponding leachates. Stock solutions of 1 g/L $Zn^{2+}$ or $As^{3+}$ were prepared by dissolving 0.44 g of $ZnSO_4·7H_2O$ or 0.13 g of $As_2O_3$ in 100 mL of ultrapure water. Batch tests were conducted similarly to the procedure described in Section 2.2, with the modification that $ZnSO_4·7H_2O$ or $As_2O_3$ replaced TWP.

**2.7 Analytical methods and statistical analysis**

The levels of $NH_4^+$-N, $NO_2^-$-N, and $NO_3^-$-N were determined according to standard methods (APHA, 2005). Salinity and pH were measured using a conductivity meter (HQ14d, HACH, Loveland, USA). Protein concentrations were quantified using a Bradford protein assay kit (BioWorks Biologicals, Shanghai. China). EPFRs were detected using an electron paramagnetic resonance spectrometer (EPR, EMXplus-6/1, Bruker, Rheinstetten, Germany). Nicotinamide adenine dinucleotide (NADH) content and electron transport chain (ETC) activity were analyzed following our previous study (Ye et al., 2022).

The statistical significance of the experimental variables was assessed by ANOVA and LSD post hoc tests with SPSS 17.0 (SPSS Inc., IL, USA). Differences were considered significant when the p-value was less than 0.05. All tests were conducted in triplicate, and data are presented as mean ± standard deviation.

## 3. Result

### 3.1 Additives released from TWP into the pore water

Multiple heavy metals and organic compounds were detected in P-TWPL and A-TWPL, and photoaging generally promotes the release of leachable additives from TWPL (Table 1). Zn and Mn were the primary heavy metals detected in P-TWPL and A-TWPL, with concentrations of 185,986 ± 7,732 μg/kg and 276,231 ± 5,854 μg/kg for Zn, and 259.5 ± 25.8 μg/kg and 315.7 ± 32.9 μg/kg for Mn. Trace amounts of Pb, Cr, Cu, and As were also found, ranging from 19.8 ± 0.9 μg/kg - 42.3 ± 3.4 μg/kg. In addition, the major organic compounds identified in P-TWPL and A-TWPL included N-Cyclohexylthiophthalimide (CTP), Dicyclohexylamine, Benzothiazole, 4-methylaniline, Acaclohexylamine, and Acaclohexylamine, methylaniline, Acetophenone, N-(1,3-Dimethylbutyl)-N'-phenyl-p-phenylenediamine (6-PPD), and Quinone derivative of 6-PPD (6-PPD-Q), with concentrations ranging from 18.2 ± 1.5 ug/kg - 61352 ± 2186 ug/kg.

The concentrations of most heavy metals (Mn, Pb, Cr, Cu) and organic compounds (CTP, Dicyclohexylamine, Benzothiazole, Acetophenone) in the pore water remained unchanged. However, Zn and As concentrations exhibited significant increases in P-TWP, A-TWP, P-TWPL, and A-TWPL after exposure, with increments of 68.4 ± 9.3% - 296.6 ± 21.4% and 49.3 ± 7.5% - 97.7 ± 12.4% compared to the control. Notably, the pore water did not detect 4-methylaniline, 6-PPD, and 6-PPDQ released from TWP.

### 3.2 Responses of nitrogen removal rates and $N_2O$ accumulation to TWP and

**TWPL**

Exposure to TWP inhibited the nitrogen removal rates in the sediment, and this inhibitory effect was generally enhanced by photoaging (Fig. 1A, B). After exposure to P-TWP, the denitrification rate in sediments decreased by 17.1 ± 10.0 %, while the anammox rate did not change significantly. In contrast, exposure to A-TWP reduced 31.3 ± 8.3% in denitrification and 22.1 ± 13.3% in anammox rates. Additionally, exposure to P-TWPL did not significantly alter the nitrogen removal rates in the sediment, whereas exposure to A-TWPL resulted in a 25.4 ± 10.0% reduction in the denitrification rate.

Due to changes in denitrification and anammox rates following exposure to TWP and TWPL, their relative contributions to nitrogen removal varied accordingly (Fig. 1C, D). After exposure to P-TWP, A-TWP, and A-TWPL, the relative contribution of denitrification to nitrogen removal declined from 91.6 ± 2.6% in the control to 73.7 ± 5.5%, 82.4 ± 3.2%, and 84.5 ± 4.2%, respectively. Conversely, the relative contribution of anammox to nitrogen removal increased from 8.4 ± 2.6% in the control to 26.3 ± 5.5%, 17.6 ± 3.2%, and 15.5 ± 4.2%, respectively. In addition, P-TWPL exposure did not significantly alter the contributions of denitrification and anammox to nitrogen removal.

Microbial denitrification removes reactive nitrogen from estuarine and coastal ecosystems while emitting the potent greenhouse gas $N_2O$ (Wu et al., 2022). Therefore, $N_2O$ accumulation in sediments was monitored after exposure to TWP and TWPL (Fig. 1E, F). P-TWP, A-TWP, and A-TWPL increased $N_2O$ accumulation in sediments by 28.1 ± 18.7%, 43.1 ± 22.0%, and 32.3 ± 23.3%, respectively. However, P-TWPL did not significantly affect N$_2$O accumulation in the sediments.

**3.3 Responses of nitrogen removal rates and N$_2$O accumulation to Zn$^{2+}$ and As$^{3+}$**

Given the significantly elevated concentrations of Zn and As in pore water following TWP and TWPL exposure, this study conducted batch experiments to assess the potentially toxic effects of Zn (931 and 1381 ug/kg dry soil) and As (0.09 and 0.15 ug/kg dry soil) released from P-TWP and A-TWP in sediments on nitrogen removal rates and N$_2$O accumulation (Fig. 2A-C). Exposure to 931 μg/kg dry soil of Zn$^{2+}$ and 0.09 - 0.15 μg/kg dry soil of As$^{3+}$ minimally impacted denitrification and anammox rates. However, 1381 μg/kg dry soil of Zn$^{2+}$ significantly reduced the denitrification rate by 17.7 ± 10.2%. Similarly, while exposure to 931 μg/kg dry soil of Zn$^{2+}$ and 0.09 - 0.15 μg/kg dry soil of As$^{3+}$ did not cause significant affect N$_2$O accumulation, 1381 μg/kg dry soil of Zn$^{2+}$ led to a 7.1 ± 6.2% increase in N$_2$O accumulation.

**3.4 Responses of gene abundance, enzyme activity, NADH content, and ETC activity to TWP and TWPL**

Exposure to TWP and TWPL significantly affected the abundance of functional genes related to denitrification and anammox in the sediments (Fig. 3). Specifically, exposure to P-TWP led to decreases in *narG* and *nirK* gene abundances of 17.5 ± 11.4% and 16.4 ± 15.7%, compared to the control. Conversely, exposure to P-TWPL did not alter *narG* gene abundance but led to an 18.8 ± 16.7% decrease in *nirK* gene abundance. More pronounced decreases were observed with A-TWP and A-TWPL, where *narG* and *nirK* gene abundances were decreased by 32.2 ± 9.4% - 37.4 ± 10.3% and 32.4 ± 27.4% - 35.0 ± 21.1%. Additionally, *hzo* gene abundance decreased by 18.7 ± 13.8% and 13.2 ± 12.3% after exposure to A-TWP and A-TWPL. Notably, *nirS* gene abundance even increased by 15.9 ± 6.2% - 19.8 ± 9.7% after exposure to P-TWP and P-TWPL. Meanwhile, there were no significant changes in *norB* and *nosZ* gene abundances across all treatment groups.

Enzyme activities and gene abundances related to denitrification and anammox in the sediment exhibited different response patterns (Fig. 4B). Exposure to P-TWP decreased NAR and NOS activities by 20.0 ± 11.7% and 23.5 ± 8.7%, compared to the control, with no significant changes observed under P-TWPL exposure. More pronounced decreases occurred with A-TWP and A-TWPL, where NAR and NOS activities decreased by 27.0 ± 12.7% - 33.0 ± 9.6% and 26.7 ± 6.9% - 30.6 ± 6.0%, and HZO activity dropped by 17.8 ± 11.0% - 27.6 ± 11.8%. Notably, NIR activity remained stable across all treatments. Additionally, while NADH content was unchanged, ETC activity decreased by 10.8 ± 9.1 %, 23.3 ± 7.1 %, and 14.3 ± 7.0 % after exposures to P-TWP, A-TWP, and A-TWPL.

**3.5 Responses of microbial community structures to TWP and TWPL**

Changes in microbial community structure and diversity after exposure to TWP and TWPL were determined by Illumina high-throughput sequencing (Fig. 5). The five most abundant phyla in the sediment included *Proteobacteria* (25.0 ± 0.7 % - 28.3 ± 1.1 %), *Desulfobacterota* (8.3 ± 0.5 % - 13.6 ± 0.6 %), *Acidobacteriota* (7.3 ± 1.3% - 10.2 ± 1.1 %), *Chloroflexi* (7.7 ± 1.9 % - 8.7 ± 2.7 %), and *Bacteroidota* (5.0 ± 0.7 % - 9.4 ± 2.6 %) (Fig. 5A). The relative abundances of the phyla *Nitrospinota*,

*Dadabacteria*, and *Zixibacteria* significantly increased under exposure to P-TWP and P-TWPL, with increments ranging from 102.8 ± 26.6% - 592.6 ± 130.0%, 98.3 ± 36.8% - 113.4 ± 36.8%, and 83.1 ± 48.1% - 138.5 ± 51.3%. In contrast, these phyla decreased under A-TWP and A-TWPL exposure, showing reductions from 36.2 ± 16.4% - 59.3 ± 8.9%, 17.5 ± 10.8% - 41.5 ± 14.1%, and 27.0 ± 13.9% - 30.4 ± 18.5%. Meanwhile, the relative abundances of the phyla *Spirochaetota*, *Bacteroidota*, and *Deinococcota* decreased under P-TWP and P-TWPL from 9.7 ± 8.9% to 51.9 ± 9.5%, 29.1 ± 13.8% to 34.9 ± 10.1%, and 38.9 ± 24.5% to 44.9 ± 16.4%, but increased under A-TWP and A-TWPL from 10.9 ± 9.1% to 36.2 ± 21.5%, 4.8 ± 2.7% to 22.9 ± 14.9%, and 18.2 ± 11.8% to 21.8 ± 13.5% (Fig. 5B).

Changes in the relative abundance of bacteria associated with denitrification and anammox in the sediment are shown in Fig. 5C. The relative abundances of the DNB genera *Achromobacter*, *Desulfobacter*, and the order *Bacteroidales* decreased by 23.5 ± 9.3% - 50.1 ± 9.8%, 18.2 ± 15.3% - 30.0 ± 18.1%, and 11.7 ± 8.3% - 58.1 ± 15.1%, after exposure to P-TWP, P-TWPL, A-TWP, and A-TWPL. In contrast, the relative abundance of the DNB genus *Pseudomonas* increased by 13.1 ± 9.4% - 57.0 ± 35.4% following these treatments. For the AnAOB, the relative abundance of order *Actinomarinales* decreased by 26.2 ± 15.8% - 33.8 ± 22.0% after exposure to P-TWP, P-TWPL, A-TWP, and A-TWPL. Conversely, the relative abundance of the genus *Candidatus kuenenia* increased by 285.7 ± 113.5 % - 378.6 ± 166.8% after exposure to P-TWP and P-TWPL.

## 4. Discussion

**4.1 Effects of P-TWP and A-TWP on sediment nitrogen removal rates**

This study focuses on microbial-driven denitrification and anammox in sediments, as these processes are the primary pathways for nitrogen removal in estuarine ecosystems (Zhang et al., 2022; Zhou et al., 2014). After exposure to P-TWP and A-TWP, denitrification rates in sediments decreased by 17.1 ± 10.0% and 31.3 ± 8.3% (Fig. 1A), indicating that TWP exposure significantly inhibits denitrification in sediments, and photoaging further exacerbates this inhibitory effect. Some researchers contend that the denitrification rate in environmental media is directly related to the denitrification functional gene abundances and enzyme activities (Xiao et al., 2021; Ye et al., 2023). This study found that exposure to P-TWP and A-TWP reduced *narG* and *nirK* gene abundances by 12.5 ± 5.4% - 19.8 ± 5.7% and 14.1 ± 11.5% - 24.4 ± 15.7% (Fig. 3A and C), and concurrently lowered NAR enzyme activity from 25.1 ± 19.2% - 49.5 ± 21.5% (Fig. 4B). These results imply a direct link between the suppression of gene and enzyme activities involved in denitrification and the diminished denitrification rate in sediments affected by TWP. Notably, *narG* gene abundance and NAR enzyme activity were 16.7 ± 12.0% and 19.6 ± 16.3% lower after A-TWP exposure than P-TWP exposure (Fig. 4B). This suggests that photoaging exacerbates the suppression of denitrification genes and enzyme activities induced by TWP, thereby intensifying the inhibition of denitrification rate. Similar reductions in *narG* gene abundance and NAR activity that led to inhibited denitrification rate were also observed in soils exposed to 25 mg/kg chlorothalonil (Su et al., 2019a). Additionally, ETC activity significantly decreased by 10.8 ± 9.1% to 23.3 ± 7.1% after P-TWP and A-TWP exposure (Fig. 4C and D), indicating that TWP inhibits the electron transport process and that photoaging intensifies this adverse effect. Previous studies have suggested that a reduction in electron transfer efficiency can interfere with the ability of denitrification enzymes to acquire electrons, thereby inhibiting the denitrification process (Su et al., 2019b). Reduced denitrification rate and NAR activity associated with ETC inhibition have been documented in suspended sediment systems exposed to 10 mg/L Ag NPs (Liu et al., 2020). Therefore, TWP-related suppression of *narG* and *nirK* gene abundances and NAR activity underlies sediment denitrification reduction, with photoaging amplifying these effects.

The anammox process showed differing responses to P-TWP and A-TWP exposures. While P-TWP exposure did not alter anammox rate in sediments, A-TWP exposure led to a significant reduction of 22.1 ± 13.3% (Fig. 1B). Consistent with this study, previous research has shown that environmentally relevant concentrations of polyvinyl chloride MPs do not significantly affect the anammox rate in soil and anaerobic sludge (Liu et al., 2022a; Ma et al., 2024). Unfortunately, these studies did not explore the changes in the biotoxicity of MPs induced by photoaging. After exposure to A-TWP, significant reductions were observed in the *hzo* gene abundance, HZO activity, and AnAOB genus *Candidatus Kuenenia* abundance by 18.7 ± 13.8%, 17.8 ± 11.0%, and 21.8 ± 15.4% compared to control (Fig. 3, 4B, and 5C). These reductions may explain the suppression of the anammox rate induced by A-TWP, as they are considered critical indicators of anammox activity. Anammox activity suppression due to decreased *hzo* gene abundance, HZO activity, and AnAOB genus

*Candidatus Kuenenia* abundance has been observed in freshwater sediments exposed to 50 mg/L $Ag^+$ and anaerobic sludge stressed with 20 mg/L sulfide (Dai et al., 2021; Xu et al., 2023). Therefore, P-TWP does not impact the anammox process in sediments, while A-TWP reduces the anammox rate by suppressing hzo abundance, HZO enzyme activity, and AnAOB genus *Candidatus Kuenenia* abundance.

Photoaging may alter the physicochemical properties of TWP, thus affecting its biotoxic effects (Lv et al., 2024). In this study, the surface roughness of A-TWP was significantly higher than that of P-TWP (Fig. S1C and D), potentially enhancing the particulate toxicity of TWP. Due to changes in surface roughness, zeta potential, and elemental composition, A-TWPs exhibited higher toxicity than P-TWPs towards *Bacillus subtilis*, *Halioglobus lutimaris*, and *Vibrio fischeri* (Kim et al., 2022; Liu et al., 2022b; Lv et al., 2024). Moreover, EPFR production during photoaging is considered one of the mechanisms enhancing the biotoxicity of A-TWPs (Liu et al., 2021). This study found EPFRs concentration in A-TWP was $4.2 \times 10^{17}$ spins/g, significantly higher than in P-TWP at $1.0 \times 10^{17}$ spins/g (Fig. 2D), which also could further exacerbate its impact on nitrogen removal processes in sediments. The EPFRs generated by photoaging may further induce the generation of reactive oxygen species, thereby damaging microbial cell membranes, inhibiting the activity of key functional enzymes, and disrupting critical microbial metabolic activities, including denitrification and anammox (Liu et al., 2021). Previous studies have also observed reductions in denitrification and anammox activities caused by reactive oxygen species generation (Huang et al., 2023; Wu et al., 2020). Therefore, changes in the physicochemical properties and EPFRs produced may partially account for the exacerbated decline in denitrification and anammox rates induced by A-TWP.

**4.2 Effects of P-TWP and A-TWPs on sediment N$_2$O accumulation**

Exposure to P-TWP and A-TWP both promoted N$_2$O accumulation in sediments. N$_2$O accumulation in sediments involves both production and consumption processes. N$_2$O production is mediated by NIR and NOR, encoded by *nirS*/*nirK* and *norB* genes, respectively, while its consumption is facilitated by NOS, encoded by *nosZ* gene (Wang et al., 2021). A previous study attributed the increased N$_2$O accumulation in estuarine sediments under polylactic acid MP stress to an increase in the *nirS*/*nosZ* abundance ratio (Chen et al., 2022). Similarly, estuarine sediments exposed to polyethylene MP showed a positive correlation between the N$_2$O accumulation and the (*nirS*+*nirK*)/*nosZ* gene abundance ratio (Huang et al., 2024). However, in this study, N$_2$O accumulation did not significantly correlate with the gene abundance ratios of (*nirS* + *nirK*)/*nosZ* and *norB*/*nosZ* (P > 0.05) but did show a significant positive correlation with the activity ratios of NIR/NOS and NOR/NOS (P < 0.05, Fig. 6). This suggests that N$_2$O accumulation after TWP exposure in sediments may depend on the ratios of enzyme activities involved in N$_2$O production and consumption rather than on gene abundance ratios. Microbial denitrification is an enzymatic biochemical process in which the activity of denitrifying enzymes directly regulates biochemical reductions (Ye et al., 2022). In contrast, the abundance of denitrification functional genes merely reflects the potential capacity for microbial metabolic functions at the genetic level (Ye et al., 2021). Previous studies have confirmed that N$_2$O accumulation in river sediment is predominantly governed by NOS activity rather than nosZ gene abundance (Su et al., 2019b). Moreover, this study observed significantly lower $N_2O$ accumulation in sediments exposed to A-TWP compared to P-TWP (Fig. 1E). Previous research has demonstrated that $N_2O$ is primarily produced through denitrification processes in sediments, including nitrifier denitrification and heterotrophic denitrification (Chen et al., 2022). Similar reductions in $N_2O$ accumulation due to denitrification inhibition under copper oxide nanoparticle stress have been noted in soil (Zhao et al., 2020). Therefore, the lower degree of $N_2O$ accumulation observed with A-TWP compared to P-TWP can be attributed to A-TWP causing a more severe inhibition of denitrification rate.

**4.3 Biotoxicity contribution of TWPL to TWP and potentially toxic components**

P-TWPL did not significantly affect the nitrogen removal rates and $N_2O$ accumulation in sediments, while A-TWPL led to a 25.4 ± 10.0% decrease in denitrification rate and a 32.3 ± 23.3% increase in $N_2O$ accumulation (Fig. 1). These findings indicate that A-TWPL exhibits higher biotoxicity compared to P-TWPL, and the adverse impacts of A-TWP on denitrification and $N_2O$ accumulation partly due to the toxic substances release from A-TWPL. Exposure to A-TWPL resulted in a decrease in NAR activity by 33.0 ± 9.6% and an increase in the NOR/NOS activity ratio by 27.9 ± 18.8% (Fig. 4B), which seems to explain its denitrification rate inhibition and $N_2O$ accumulation promotion. Notable, the anammox rate remained unchanged after exposure to either P-TWPL or A-TWPL, suggesting that the inhibition of anammox activity attributed to A-TWP likely results from alterations in TWP's physicochemical properties or the formation of EPFRs induced by photoaging, rather than from direct emanation from A-TWPL.

Despite no significant changes in the concentrations of six organic compounds in the pore water following exposure to P-TWPL and A-TWPL, the concentrations of Zn and As significantly increased, rising by 106.7 ± 15.5%, 296.6 ± 28.7% and 73.2 ± 44.0%, 97.7 ± 49.4% respectively (Table 1). This suggests that photoaging facilitated the release of Zn and As from TWPL, with the augmented concentrations of these elements likely playing a pivotal role in the observed reduced denitrification rate and increased $N_2O$ accumulation in sediments caused by A-TWPL.

To determine the potential impacts of Zn and As released from TWP on nitrogen removal rates and $N_2O$ accumulation in sediments, this study conducted batch experiments to evaluate the effects of the theoretical concentrations of $Zn^{2+}$ and $As^{3+}$ in sediments, following exposure to P-TWPL and A-TWPL, on denitrification rate, anammox rate, and $N_2O$ accumulation (Fig. 2). Theoretical concentrations of $As^{3+}$ in P-TWPL and A-TWPL (0.09 and 0.15 µg/kg dry soil) did not significantly affect nitrogen removal rates or $N_2O$ accumulation in sediments, suggesting that As is not a primary toxicant in TWPL. However, the addition of $Zn^{2+}$ at the theoretical concentration in A-TWPL (1381 µg/kg dry soil) resulted in a 17.7 ± 10.2% reduction in the denitrification rate and a 7.1 ± 6.2% increase in $N_2O$ accumulation. This aligns with the phenomena observed after A-TWP exposure, suggesting that Zn may partly contribute to TWPL's biotoxicity. Previous studies have found that $Zn^{2+}$ at a concentration of 0.45 mg/L inhibits the denitrification process in activated sludge by suppressing NAR enzyme activity (Zheng et al., 2011). Notably, $Zn^{2+}$ concentrations in P-TWPL and A-TWPL (931 and 1381 µg/kg dry soil) did not significantly affect the anammox rate in sediments, which may be attributed to the relatively high metal tolerance of anammox bacteria (Ma et al., 2020). Additionally, previous studies have reported that 4-methylaniline, 6-PPD, and 6-PPDQ are TWP's primary potentially toxic organic compounds. However, in this study, these substances were nearly undetectable in pore water (<1 µg/L), likely due to adsorption by sediments and microbial degradation (Arndt et al., 2013; Miranda et al., 2021). Therefore, the observed decrease in denitrification rate and increase in $N_2O$ accumulation induced by A-TWP can be partially attributed to the release of $Zn^{2+}$ from A-TWPL.

**4.4 Effects of TWPL on microbial community structures**

Considering that additives may play a significant role in the impact of TWP on microbial community functions in estuarine sediments, this study compared the differences in microbial community structure after exposure to TWP and TWPL. PCoA1 and PCoA2 explained over 57% of the variation in microbial community structure, effectively revealing the major differences between treatments (Fig. 7A). The differences in the sediment microbial communities between P-TWP and P-TWPL treatments or between A-TWP and A-TWPL treatments were minor, suggesting that TWPL may be the primary factor driving changes in the sediment microbial community structure. The PERMANOVA test also confirmed these findings, showing no significant differences in microbial community structure between the P-TWP/A-TWP and P-TWPL/A-TWPL treatments (Table S3, $P > 0.05$). Furthermore, the Shannon index (α-diversity indicator) showed no significant differences between the P-TWP/A-TWP and their leachate treatments (Fig. 7B), further supporting the notion that leachate-induced effects are the primary drivers of microbial structure changes in estuarine sediments. Previous studies have also confirmed that TWPL is the main factor behind the shifts in microbial community structure caused by TWP (Ding et al., 2022). These structural changes may indirectly impact microbial functions and diversity, potentially reducing nitrogen removal rates and increasing $N_2O$ accumulation from sediments.

Spearman correlations were analyzed between concentrations of heavy metals and organic compounds in pore water and the relative abundances of DNB and AnAOB to identify key toxicants affecting nitrogen removal microbes in sediments (Fig. 7C). Zn concentration was negatively correlated with the relative abundances of DNB f_*Steroidobacteraceae*, o_*S085*, g_*Achromobacter*, and g_*Anaeromyxobacter* ($P < 0.05$), consistent with findings that Zn release significantly impacts denitrification rate and $N_2O$ accumulation (Fig. 2). Additionally, negative correlations were observed between Zn, Mn, or As concentrations and the relative abundances of AnAOB *Actinomarinales* or *Candidatus Kuenenia* ($P < 0.05$), indicating potential adverse effects of heavy metals on the anammox process despite no significant inhibition of anammox rate. However, Zn concentration was positively correlated with the relative abundances of several DNBs (o_*SBR1031*, g_*Pseudomonas*, and f_*Rhodobacteraceae*), likely due to the functional redundancy of the microbial community (Ye et al., 2023). Notably, no significant correlations were found between organic compound concentrations and the relative abundances of DNB and AnAOB, possibly because the organic compound levels in sediments did not reach the threshold to disrupt microbial community structure. For example, the theoretical concentration of the potentially toxic organic compound benzothiazole in sediments after A-TWPL exposure was 2.2 µg/kg, insufficient to cause significantly growth rate inhibition on typical microbes such as *Raphidocelis subcapitata* and *Dunaliella tertiolecta* (Canova et al., 2021). Thus, the additive release is the main factor driving TWP-induced changes in sediment microbial community structure and diversity, with the relative abundances of DNB and AnAOB linked to Zn, Mn, and As levels in pore water.

**4.4 Environmental implications**

Estuarine ecosystems play a crucial role in mitigating nitrogen flow from estuaries to the ocean, and the stability of their nitrogen removal function is expected to be affected by the TWP. However, the effects of TWP and its leachable additives on key nitrogen removal processes and $N_2O$ accumulation in estuarine sediments, particularly the biotoxic changes induced by photoaging, have not received sufficient attention. This study suggests that photoaging may exacerbate the negative impact of TWP on key nitrogen removal processes and $N_2O$ accumulation in estuarine sediments by facilitating heavy metal release, increasing particle toxicity, and forming EPFRs. Therefore, it is critical to consider the biotoxicity changes caused by photoaging when assessing the ecological effects and risks of TWP in estuarine ecosystems. To this end, we recommend incorporating TWP into surface water quality standards and monitoring its photoaging process, including the release of toxic substances and the formation of toxic by-products, to mitigate the adverse effects of photoaging on TWP.

## 5. Conclusions

This study comprehensively compared the effects and potential toxicity mechanisms of P-TWP and A-TWP and their leachates on key nitrogen removal processes and $N_2O$ accumulation in estuarine sediments and explored the potential biotoxic substances in TWPL. The main conclusions of this study are:

1) P-TWP inhibited the denitrification rate and promoted $N_2O$ accumulation without significantly affecting the anammox rate.

2) A-TWP intensified the denitrification rate inhibition by exacerbating the suppression of *narG* gene abundance and NAR activity, and reduced anammox rate by decreasing *hzo* gene abundance, HZO activity, and *Candidatus Kuenenia* abundance.

3) Lower $N_2O$ accumulation after exposure to A-TWP compared to P-TWP is due to more pronounced denitrification inhibition by A-TWP, with NIR/NOS and NOR/NOS activity ratios crucially regulating sediment $N_2O$ accumulation.

4) The denitrification rate inhibition and $N_2O$ accumulation promotion induced by A-TWP is partly due to the increased Zn concentration in A-TWPL, and TWP alters changes in sediment microbial community structure through release additives.

These findings provide new insights into the impacts of TWPs on key nitrogen removal processes in estuarine environments and highlight photoaging's important role in modulating TWP biotoxicity.

**Acknowledgment**

This work was supported by the National Natural Science Foundation of China (42307479) and the Fundamental Research Funds of Zhejian University of Science and Technology (2023QN017). The authors would like to thank the shiyanjia lab (www.shiyanjia.com) for the MIMS, IRMS, and EPR analysis.

**Conflicts of interest**

There are no conflicts of interest to declare.

**Table Captions**

Table 1. Heavy Metals and Organic Compounds Detected in leachate and pore water.

| Items | TWPL (μg/kg) | | Pore water (μg/L) | | | | |
|---|---|---|---|---|---|---|---|
| | P-TWPL | A-TWPL | Control | P-TWP | P-TWPL | A-TWP | A-TWPL |
| Zn | 185986 ± 7732 | 276231 ± 5854 | 113.9 ± 7.5 | 191.8 ± 5.1$^a$ | 235.4 ± 8.4$^a$ | 363.1 ± 6.8$^a$ | 451.7 ± 13.6$^a$ |
| Mn | 259.5 ± 25.8 | 315.7 ± 32.9 | 6532.7 ± 685.4 | 6385.5 ± 574.3 | 6734.1 ± 548.3 | 6981.5 ± 920.3 | 7443.2 ± 897.6 |
| Pb | 35.4 ± 1.3 | 42.6 ± 4.8 | 14.7 ± 6.1 | 13.6 ± 4.8 | 17.8 ± 5.6 | 17.7 ± 6.3 | 19.5 ± 6.7 |
| Cr | 32.8 ± 3.5 | 37.5 ± 6.1 | 3.1 ± 1.1 | 4.2 ± 1.3 | 4.8 ± 1.4 | 5.7 ± 2.2 | 5.3 ± 1.6 |
| Cu | 28.5 ± 1.2 | 42.3 ± 3.4 | 5.4 ± 1.8 | 5.9 ± 2.0 | 7.8 ± 2.4 | 6.7 ± 1.5 | 8.2 ± 2.9 |
| As | 19.8 ± 0.9 | 29.6 ± 2.9 | 21.3 ± 4.2 | 31.8 ± 5.1$^a$ | 36.9 ± 5.7$^a$ | 39.5 ± 4.8$^a$ | 42.1 ± 6.2$^a$ |
| CTP | 33872 ± 2075 | 31596 ± 3146 | 1.7 ± 0.7 | 2.1 ± 0.9 | 3.2 ± 1.3 | 2.8 ± 0.8 | 3.6 ±1.8 |
| Dicyclohexylamine | 58741 ± 1032 | 61352 ± 2186 | 1.4 ± 0.6 | 1.5 ± 0.8 | 2.7 ± 1.3 | 2.0 ± 1.1 | 2.6 ± 1.4 |
| Benzothiazole | 4489.3 ± 168.8 | 4754.7 ± 385.7 | 1.9 ± 0.9 | 1.3 ± 0.6 | 2.1 ± 1.5 | 1.8 ± 1.3 | 2.2 ± 1.1 |
| 4-methylaniline | 3786.4 ± 192.3 | 3581.7 ± 173.2 | N.D. | N.D. | N.D. | N.D. | N.D. |
| Acetophenone | 743.6 ± 73.1 | 736 ± 85.8 | 6.2 ± 2.1 | 4.3 ± 1.6 | 7.4 ± 2.5 | 5.1 ± 1.9 | 8.2 ± 3.1 |
| 6-PPD | 146.7 ± 13.2 | 136.8 ± 21.6 | N.D. | N.D. | N.D. | N.D. | N.D. |
| 6-PPD-Q | 18.2 ± 1.5 | 35.4 ± 4.5 | N.D. | N.D. | N.D. | N.D. | N.D. |

$^a$: Indicates a significant difference from the control (P < 0.05); N.D. = not detected ( < 1 ug/L).

**Figure Captions**

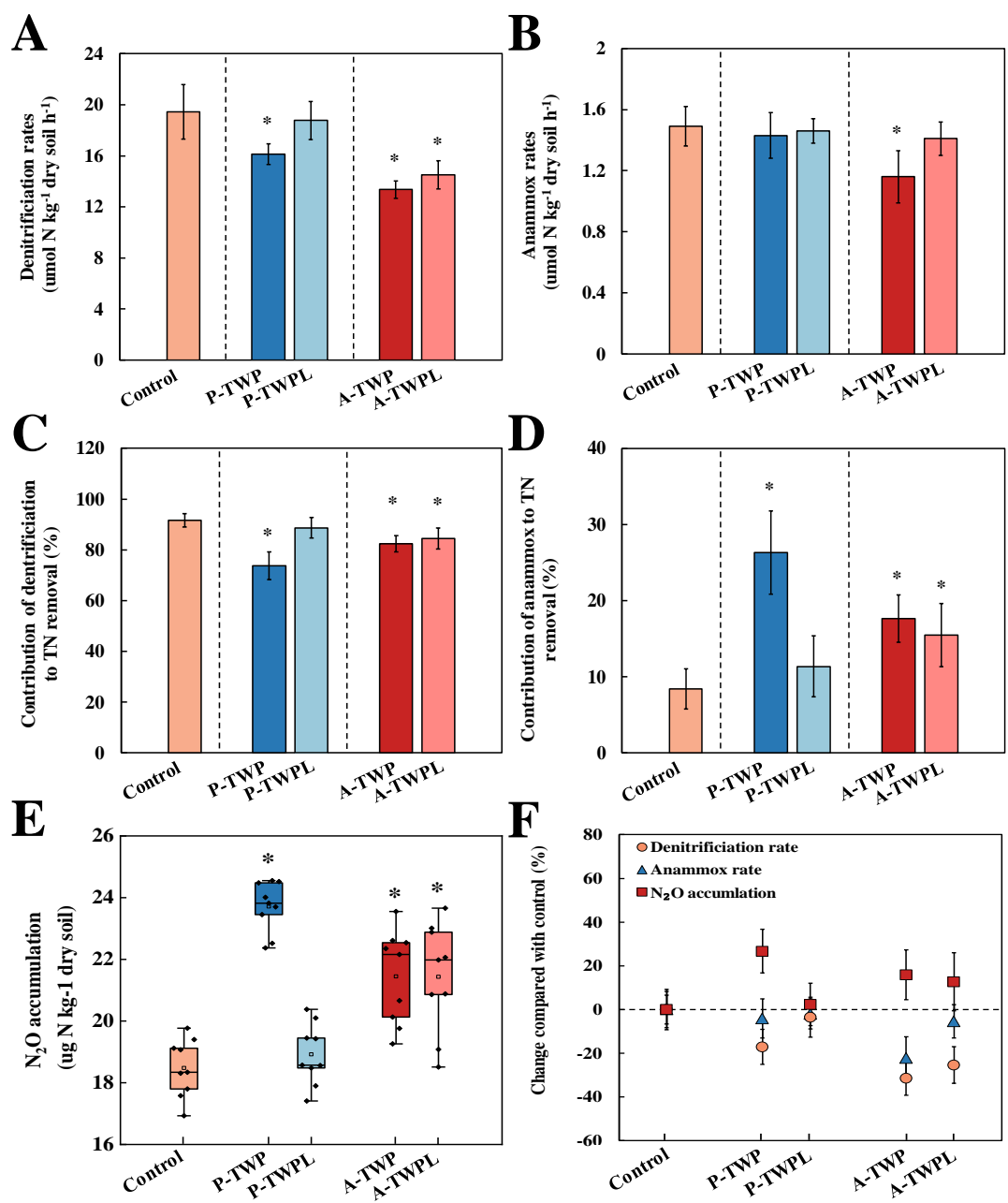

Fig. 1 Denitrification rate (A), anammox rate (B), the contribution of denitrification to nitrogen removal (C), the contribution of anammox to nitrogen removal (D), N$_2$O accumulation (E), and the changes in denitrification rate/anammox rate/N$_2$O accumulation (F) in estuarine sediments after exposure to P-TWP, A-TWP, P-TWPL, and A-TWPL. "*" denotes significant differences from the control (P < 0.05).

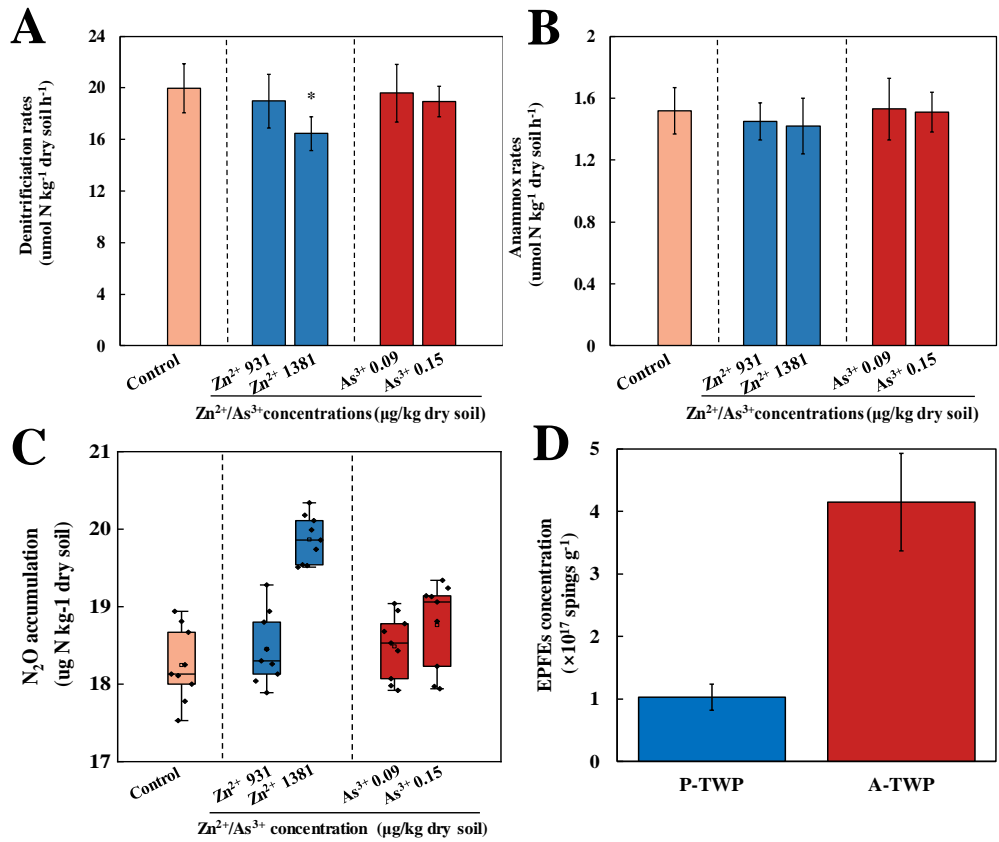

Fig. 2 Denitrification rate (A), anammox rate (B), and N$_2$O accumulation (C) in estuarine sediments after exposure to theoretical concentrations of Zn$^{2+}$ and As$^{3+}$ in P-TWPL and A-TWPL. EPFRs concentration on P-TWP and A-TWP (D). "*" denotes significant differences from the control (P < 0.05).

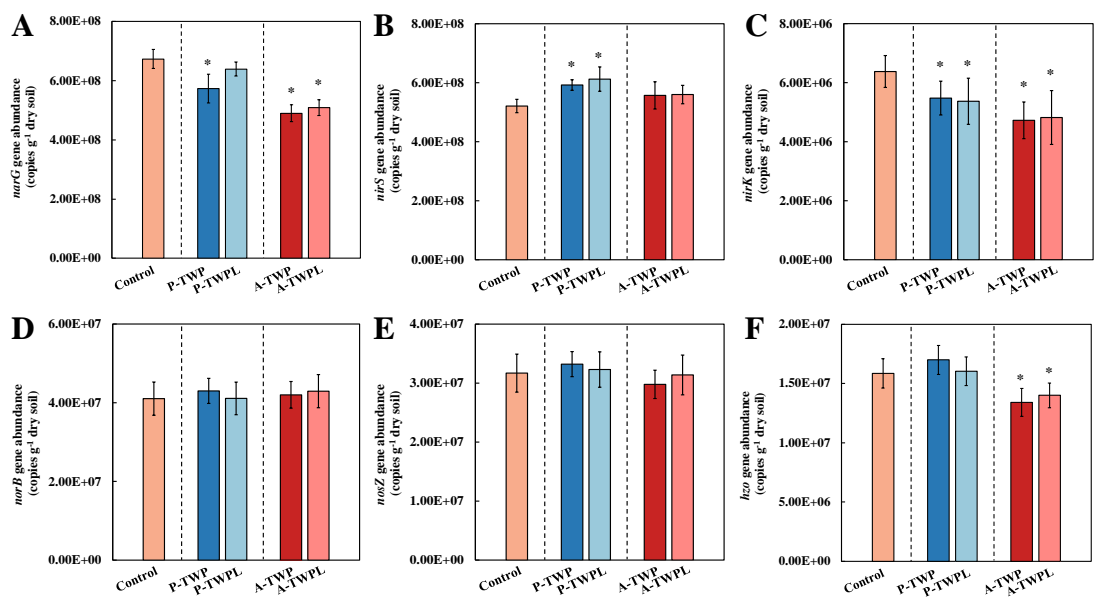

Fig. 3 Gene abundances of *narG* (A), *nirS* (B), *nirK* (C), *norB* (D), *nosZ* (E), and *hzo* (F) genes in estuarine sediments. "*" denotes significant differences from the control (P < 0.05).

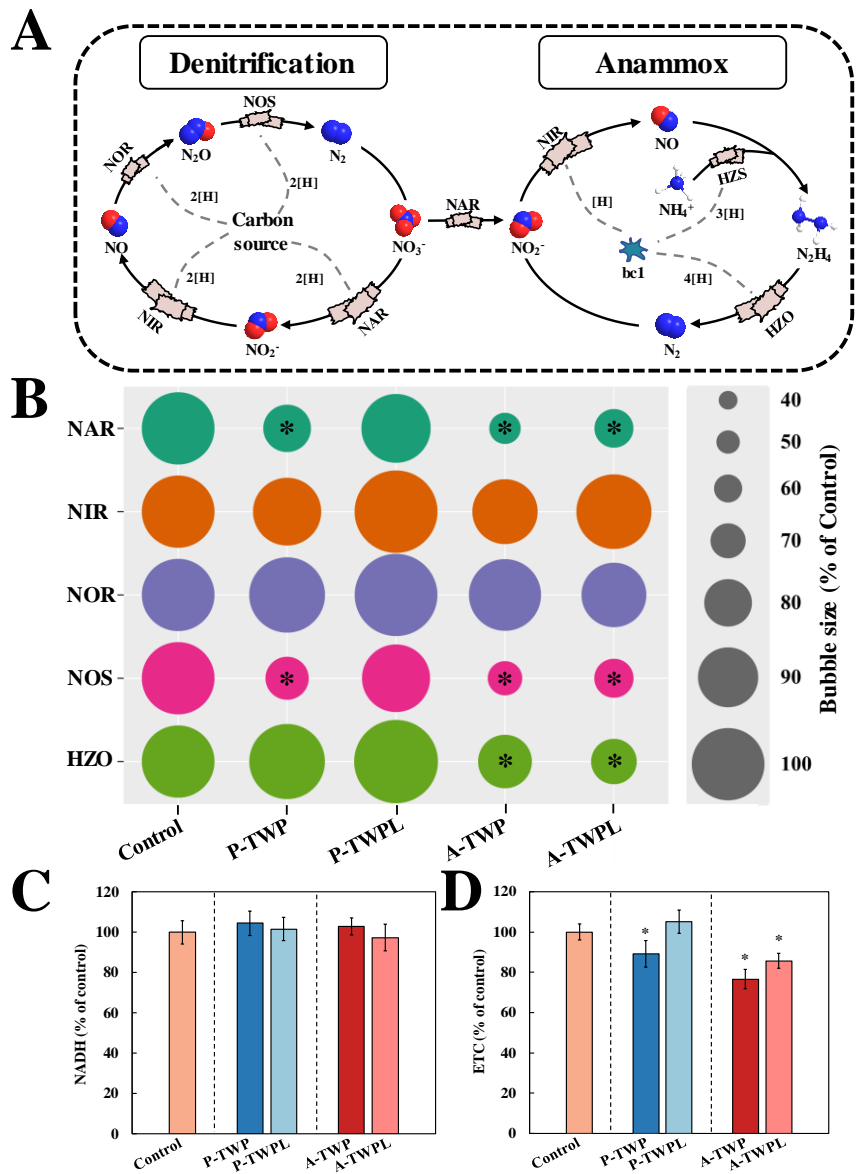

Fig. 4 Schematic diagram of denitrification and anammox processes (A). Enzyme activities of NAR, NIR, NOR, NOS, and HZO in estuarine sediments, with each activity shown in bubble plot size (B). NADH content (C) and ETC activity (D) in estuarine sediments after exposure to P-TWP, A-TWP, P-TWPL, and A-TWPL. "*" denotes significant differences from the control ($P < 0.05$).

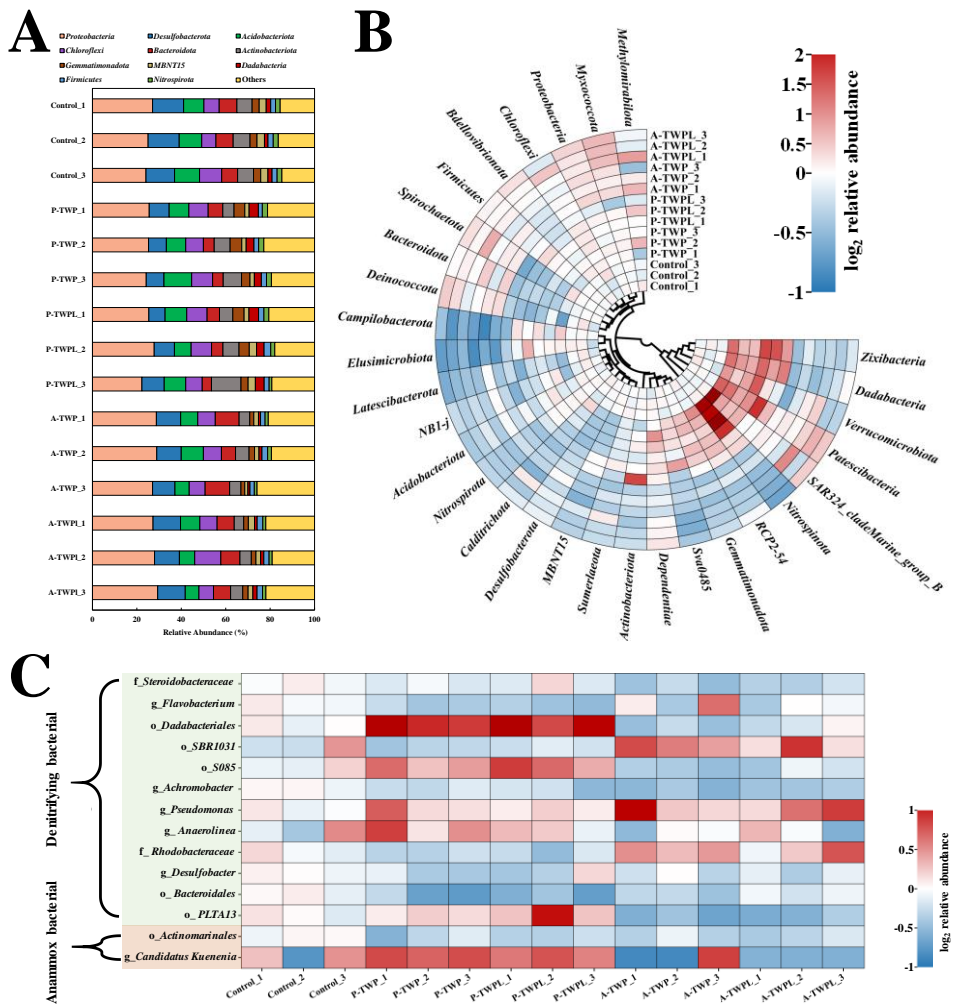

Figure 5 Relative abundance of dominant phyla (A), changes in the relative abundance of dominant phyla (B), and changes in the relative abundance of DNB and AnAOB (C) in estuarine sediments after exposure to P-TWP, A-TWP, P-TWPL, and A-TWPL.

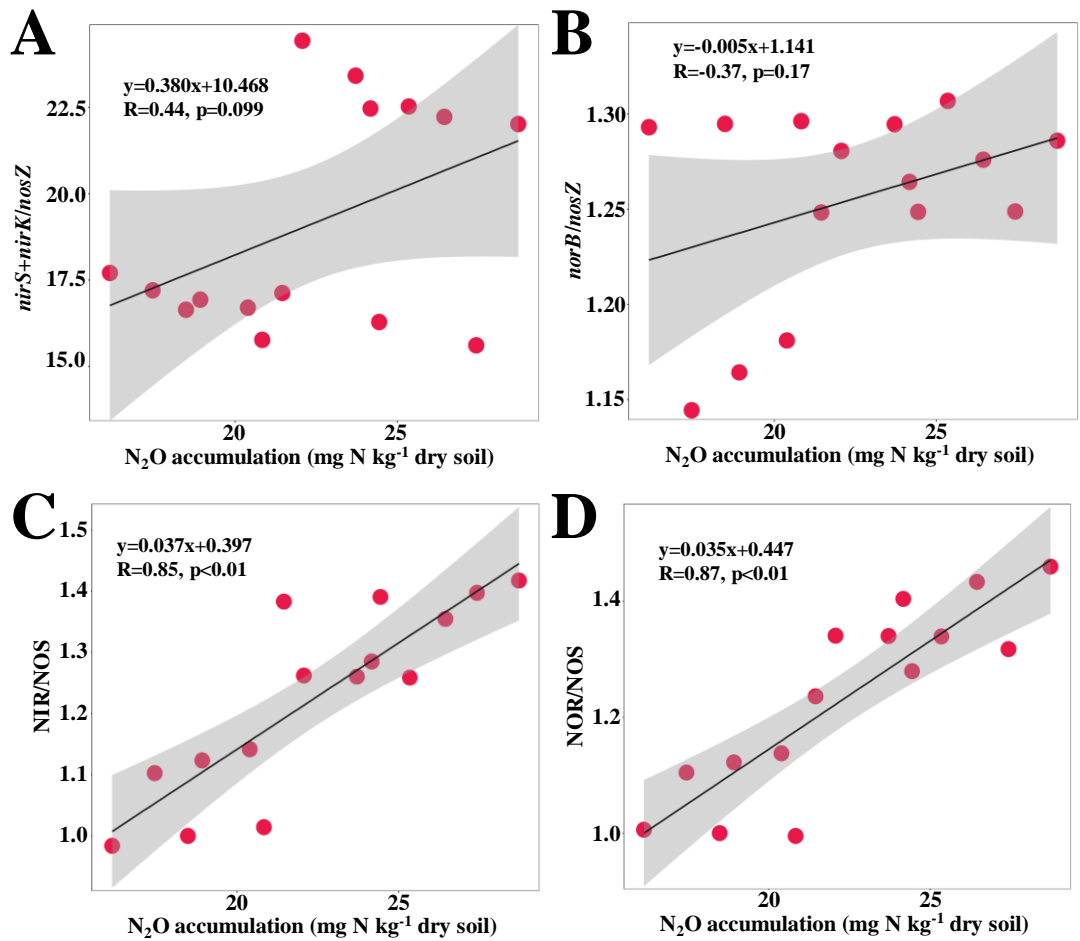

Fig. 6 Correlations between the abundance ratios of N₂O-producing genes/N₂O-consuming genes (A, B) and the activity ratios of N₂O-producing enzymes/N₂O-consuming enzymes (C, D) with N₂O accumulation.

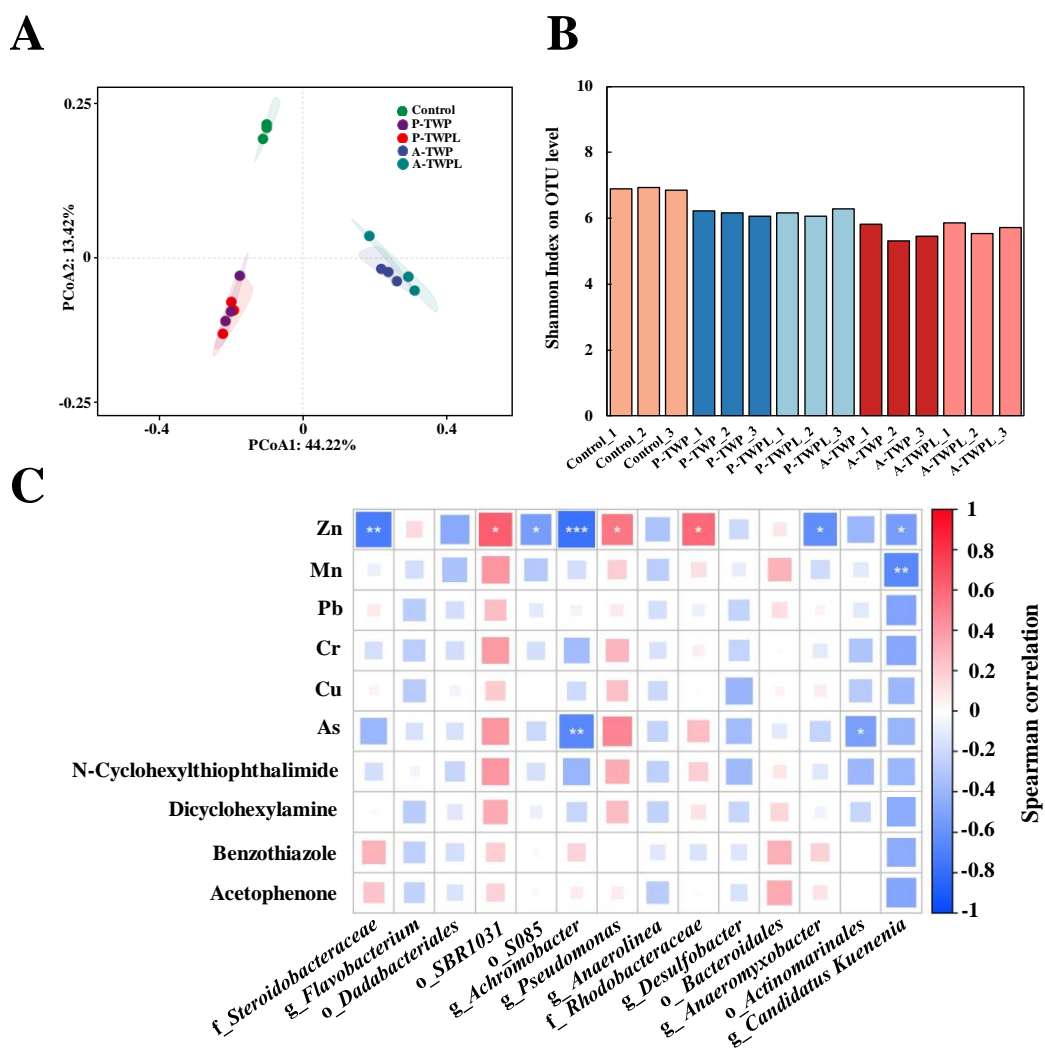

Fig. 7 PCoA of Bray-Curtis distance (A), Shannon index on OTU level (B), and Spearman's correlation of heavy metals and organic compounds with relative abundance of DNB and AnAOB (C) in estuarine sediments after exposure to P-TWP, A-TWP, P-TWPL, and A-TWPL. "*" denotes significant differences from the control ($P < 0.05$).

**TITLE**

Effects of pristine and photoaged tire wear particles and their leachable additives on key nitrogen removal processes and nitrous oxide accumulation in estuarine sediments

**AUTHOR LIST**

Jinyu Ye [a,b*], Yuan Gao [a,b], Huan Gao [c], Qingqing Zhao [a,b], Minjie Zhou [d], Xiangdong Xue [a,b*], Meng Shi [e]

a. School of Civil Engineering and Architecture, Zhejiang University of Science and Technology, Hangzhou, Zhejiang, 310023, China

b. Zhejiang-Singapore Joint Laboratory for Urban Renewal and Future City, Hangzhou, 310023, China

c. Key Laboratory of Pesticide & Chemical Biology of Ministry of Education, College of Chemistry, Central China Normal University, Wuhan, 430079, China

d. Pingyang County Aojiang River Basin Water Conservancy Project Management Center, Wenzhou, Zhejiang,325401, China

e. Center for Energy Resources Engineering, Department of Chemical and Biochemical Engineering, Technical University of Denmark, Kongens Lyngby, 2800, Denmark

* Corresponding author: Xiangdong Xue, School of Civil Engineering and Architecture, Zhejiang University of Science and Technology, No.318 Liuhe Road, Zhejiang, 310023, China; E-mail: water21cn_xxd@163.com

Table S1. Physicochemical properties of the sediments and overlying water.

| Index | Sediment | Overlying water |
| --- | --- | --- |
| pH | - | 8.05 |
| $NH_4^+$ | 7.15 mg/kg | 0.12 mg/L |
| $NO_2^-$ | 0.21 mg/kg | 0.04 mg/L |
| $NO_3^-$ | 3.13 mg/kg | 1.92 mg/L |
| Salinity | - | 11.36 |

Table S2. Primers and conditions used for the quantification of specific genes by qPCR.

| Target gene | Primer | Sequence (5' - 3') | Annealing temperature | Reference |
|---|---|---|---|---|
| *narG* | narG1960f | TAY GTS GGS CAR GAR AA | 55 | (Philippot, 2002) |
|  | narG2650r | TYT CRT ACC ABG TBG C |  |  |
| *napA* | napA v66 | TAY TTY YTN HSN AAR ATH ATG TAY GG | 50 | (Flanagan et al., 1999) |
|  | napA v67 | DAT NGG RTG CAT YTC NGC CAT RTT |  |  |
| *nirS* | Cd3aF | GTS AAC GTS AAG GAR ACS GG | 45 | (Throbäck et al., 2004) |
|  | R3cd | GAS TTC GGR TGS GTC TTG A |  |  |
| *nirK* | F1aCu | ATC ATG GTS CTG CCG CG | 57 | (Hallin and Lindgren, 1999) |
|  | R3Cu | GCC TCG ATC AGR TTG TGG TT |  |  |
| *nosZ* | NosZ 1527F | CGC TGT TCH TCG ACA GYC A | 60 | (Liu et al., 2016) |
|  | NosZ 1773R | ATR TCG ATC ARC TGB TCG TT |  |  |
| *hzo* | hzocl1F1 | TGY AAG ACY TGY CAY TGG G | 60 | (Schmid et al., 2008) |
|  | hzocl1R2 | ACT CCA GAT RTG CTG ACC |  |  |

Table S3. Differences in sediment microbial community structure between different TWP and TWPL treatments

| Treatment | $R^2$ | $P$ |
|---|---|---|
| Control VS P-TWP | 0.459 | 0.011 |
| Control VS A-TWP | 0.365 | 0.027 |
| Control VS P-TWPL | 0.488 | 0.014 |
| Control VS A-TWPL | 0.305 | 0.029 |
| P-TWP VS P-TWPL | 0.192 | 0.109 |
| P-TWP VS A-TWPL | 0.501 | 0.014 |
| P-TWP VS A-TWP | 0.478 | 0.017 |
| A-TWP VS A-TWPL | 0.126 | 0.121 |

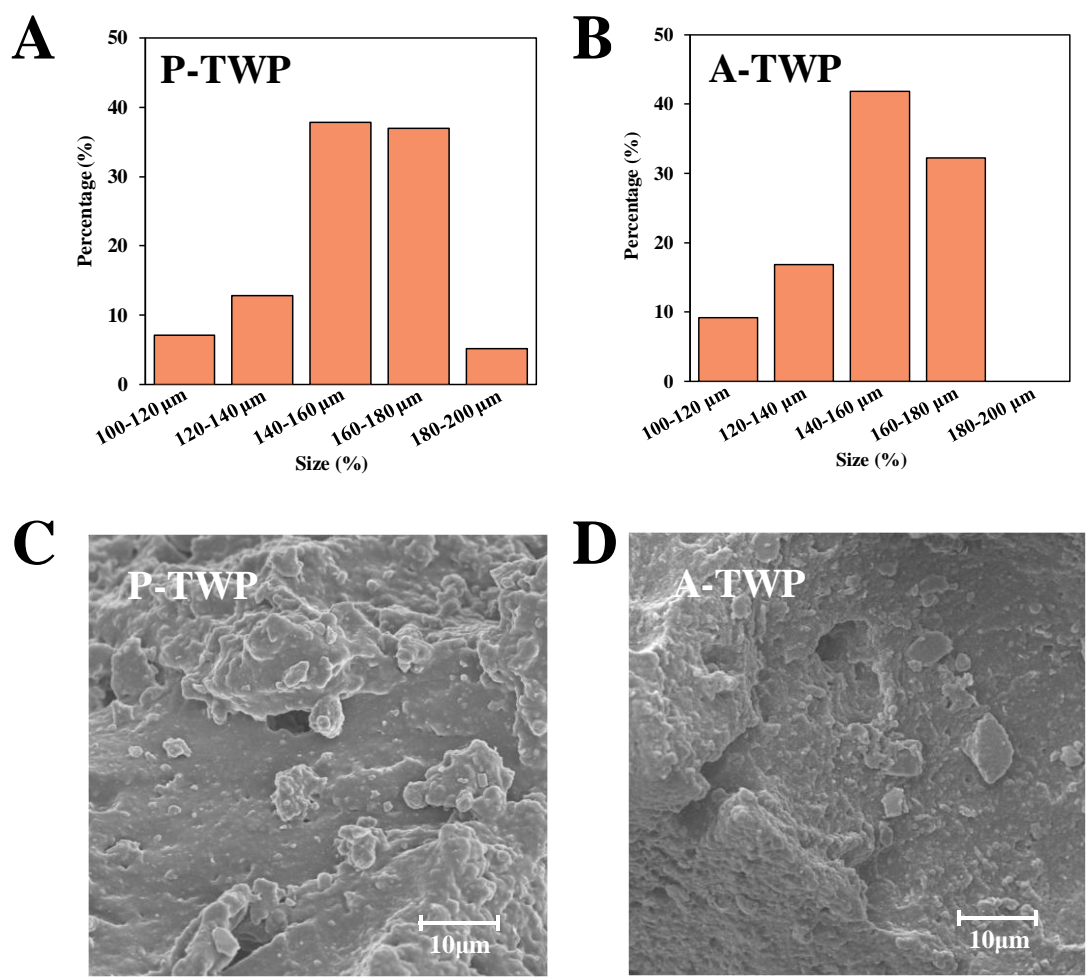

Figure S1. Particle size distribution (A, B) and surface morphology (C, D) of P-TWP and A-TWP.

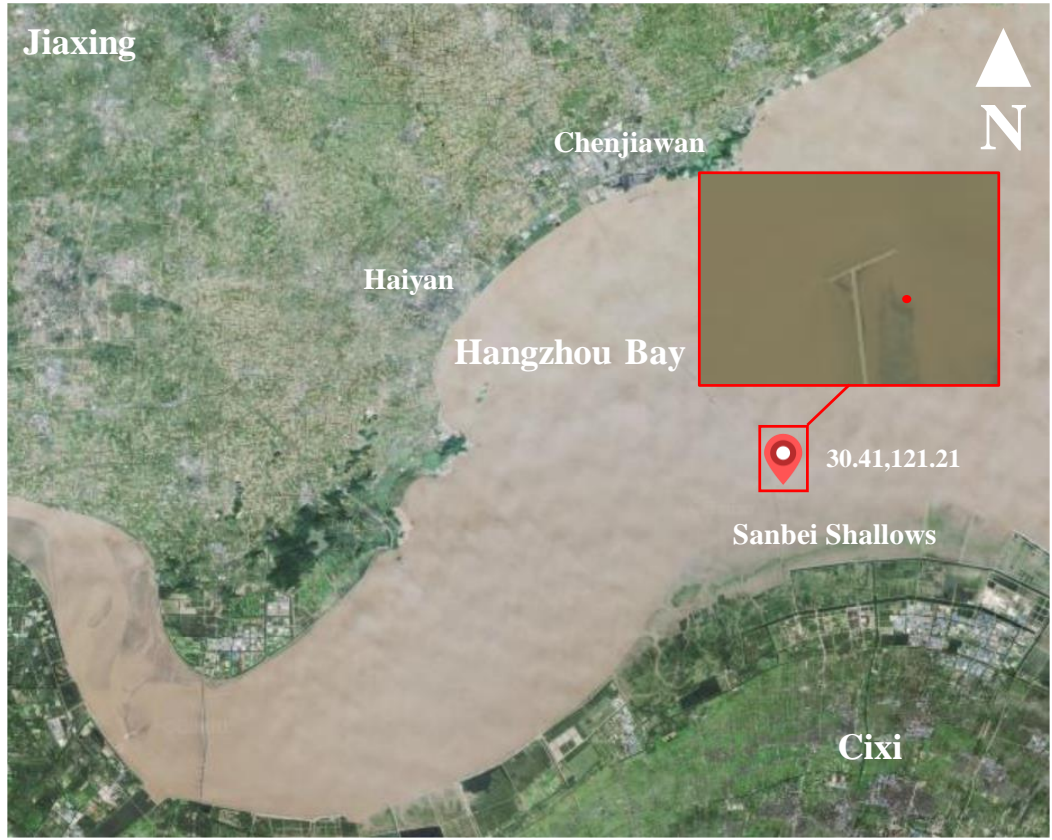

Figure S2. Map of the Hangzhou Bay showing the sample site.

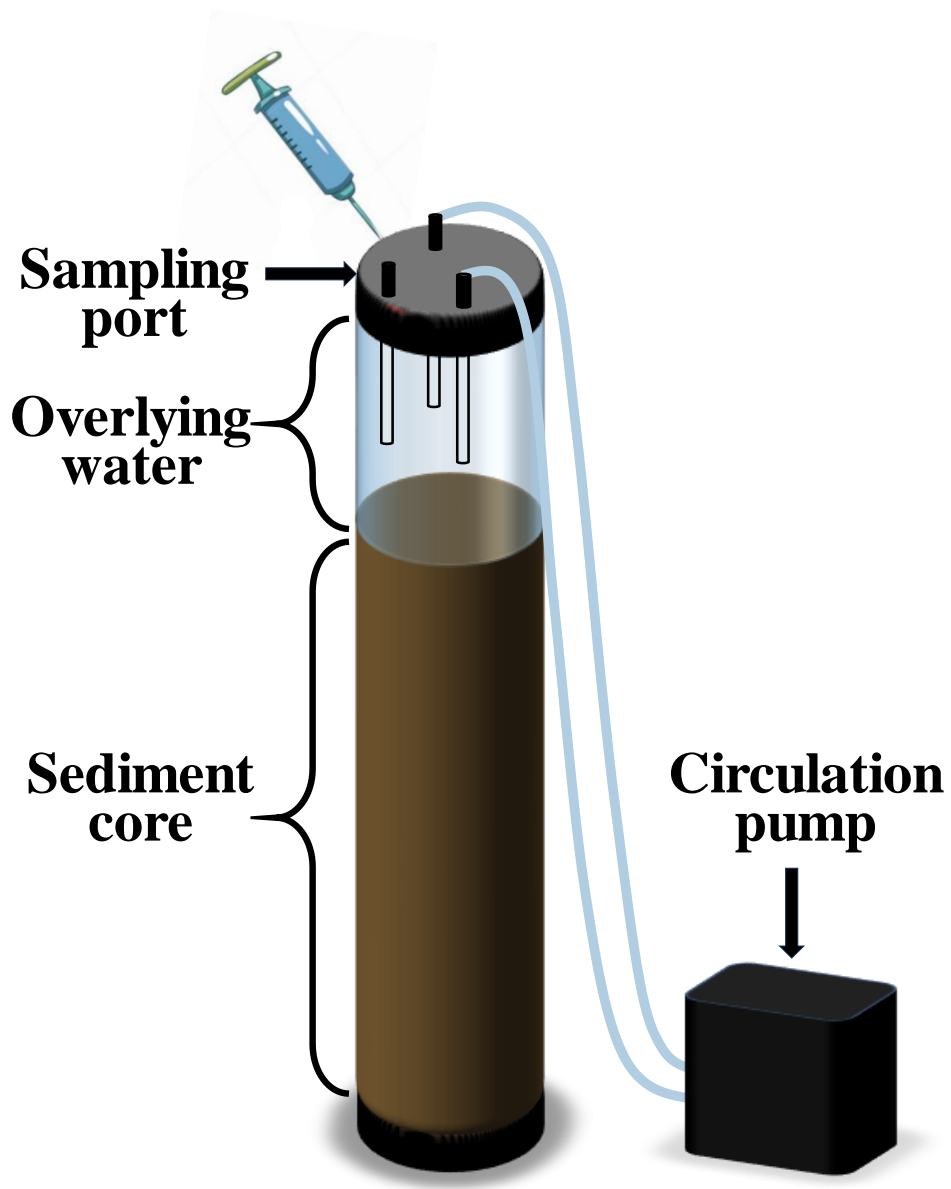

Figure S3. Schematic diagram of the sediment core.